\documentstyle[12pt]{article}
\date{}
\author{Valerii Dryuma\thanks{Work supported in part by Grant RFFI, Russia-Moldova}\\[5mm]
{\it Institute of Mathematics and Informatics, AS RM,}\\[3mm] {\it
5 Academiei str., 2028 Kishinev, Moldova},\\[3mm]{\it e-mail:
valery@dryuma.com; emental@mail.md} }
\title{On nonlinear equations associated with \\[1mm] developable, ruled and minimal surfaces}

\begin{document}
\maketitle
\date{}
\maketitle
\begin{abstract}
\ \ \ \ An examples of solutions of nonlinear differential equations associated with
 developable, ruled and minimal surfaces are constructed.
\end{abstract}


\section{ Developable surfaces}

    The equation of developable surface is defined by the condition \cite{Bysh}
    \begin{equation}\label{dr:eq1}
\left[ \begin {array}{cccc} {\frac {\partial ^{2}}{\partial {x}^{2}}}
F \left( x,y,z \right) &{\frac {\partial ^{2}}{\partial x\partial y}}F
 \left( x,y,z \right) &{\frac {\partial ^{2}}{\partial x\partial z}}F
 \left( x,y,z \right) &{\frac {\partial }{\partial x}}F \left( x,y,z
 \right) \\\noalign{\medskip}{\frac {\partial ^{2}}{\partial x
\partial y}}F \left( x,y,z \right) &{\frac {\partial ^{2}}{\partial {y
}^{2}}}F \left( x,y,z \right) &{\frac {\partial ^{2}}{\partial y
\partial z}}F \left( x,y,z \right) &{\frac {\partial }{\partial y}}F
 \left( x,y,z \right) \\\noalign{\medskip}{\frac {\partial ^{2}}{
\partial x\partial z}}F \left( x,y,z \right) &{\frac {\partial ^{2}}{
\partial y\partial z}}F \left( x,y,z \right) &{\frac {\partial ^{2}}{
\partial {z}^{2}}}F \left( x,y,z \right) &{\frac {\partial }{\partial
z}}F \left( x,y,z \right) \\\noalign{\medskip}{\frac {\partial }{
\partial x}}F \left( x,y,z \right) &{\frac {\partial }{\partial y}}F
 \left( x,y,z \right) &{\frac {\partial }{\partial z}}F \left( x,y,z
 \right) &0\end {array} \right]=0.
\end{equation}

   It is equivalent to the  second order partial differential equation
    \begin{equation}\label{dr:eq2}
- \left( {\frac {\partial ^{2}}{\partial {x}^{2}}}F \left( x,y,z
 \right)  \right)  \left( {\frac {\partial ^{2}}{\partial {y}^{2}}}F
 \left( x,y,z \right)  \right)  \left( {\frac {\partial }{\partial z}}
F \left( x,y,z \right)  \right) ^{2}+$$$$+2\, \left( {\frac {\partial ^{2}}
{\partial {x}^{2}}}F \left( x,y,z \right)  \right)  \left( {\frac {
\partial ^{2}}{\partial y\partial z}}F \left( x,y,z \right)  \right)
 \left( {\frac {\partial }{\partial y}}F \left( x,y,z \right)
 \right) {\frac {\partial }{\partial z}}F \left( x,y,z \right) -$$$$-
 \left( {\frac {\partial ^{2}}{\partial {x}^{2}}}F \left( x,y,z
 \right)  \right)  \left( {\frac {\partial }{\partial y}}F \left( x,y,
z \right)  \right) ^{2}{\frac {\partial ^{2}}{\partial {z}^{2}}}F
 \left( x,y,z \right) +$$$$+ \left( {\frac {\partial ^{2}}{\partial x
\partial y}}F \left( x,y,z \right)  \right) ^{2} \left( {\frac {
\partial }{\partial z}}F \left( x,y,z \right)  \right) ^{2}-$$$$-2\,
 \left( {\frac {\partial ^{2}}{\partial x\partial y}}F \left( x,y,z
 \right)  \right)  \left( {\frac {\partial ^{2}}{\partial y\partial z}
}F \left( x,y,z \right)  \right)  \left( {\frac {\partial }{\partial x
}}F \left( x,y,z \right)  \right) {\frac {\partial }{\partial z}}F
 \left( x,y,z \right) -$$$$-2\, \left( {\frac {\partial ^{2}}{\partial x
\partial y}}F \left( x,y,z \right)  \right)  \left( {\frac {\partial }
{\partial y}}F \left( x,y,z \right)  \right)  \left( {\frac {\partial
^{2}}{\partial x\partial z}}F \left( x,y,z \right)  \right) {\frac {
\partial }{\partial z}}F \left( x,y,z \right) +$$$$+2\, \left( {\frac {
\partial ^{2}}{\partial x\partial y}}F \left( x,y,z \right)  \right)
 \left( {\frac {\partial }{\partial y}}F \left( x,y,z \right)
 \right)  \left( {\frac {\partial }{\partial x}}F \left( x,y,z
 \right)  \right) {\frac {\partial ^{2}}{\partial {z}^{2}}}F \left( x,
y,z \right) +$$$$+2\, \left( {\frac {\partial ^{2}}{\partial x\partial z}}F
 \left( x,y,z \right)  \right)  \left( {\frac {\partial ^{2}}{
\partial {y}^{2}}}F \left( x,y,z \right)  \right)  \left( {\frac {
\partial }{\partial x}}F \left( x,y,z \right)  \right) {\frac {
\partial }{\partial z}}F \left( x,y,z \right) +$$$$+ \left( {\frac {
\partial ^{2}}{\partial x\partial z}}F \left( x,y,z \right)  \right) ^
{2} \left( {\frac {\partial }{\partial y}}F \left( x,y,z \right)
 \right) ^{2}-$$$$-2\, \left( {\frac {\partial ^{2}}{\partial x\partial z}}
F \left( x,y,z \right)  \right)  \left( {\frac {\partial }{\partial y}
}F \left( x,y,z \right)  \right)  \left( {\frac {\partial }{\partial x
}}F \left( x,y,z \right)  \right) {\frac {\partial ^{2}}{\partial y
\partial z}}F \left( x,y,z \right) -$$$$- \left( {\frac {\partial ^{2}}{
\partial {y}^{2}}}F \left( x,y,z \right)  \right)  \left( {\frac {
\partial }{\partial x}}F \left( x,y,z \right)  \right) ^{2}{\frac {
\partial ^{2}}{\partial {z}^{2}}}F \left( x,y,z \right) + $$$$+\left( {
\frac {\partial }{\partial x}}F \left( x,y,z \right)  \right) ^{2}
 \left( {\frac {\partial ^{2}}{\partial y\partial z}}F \left( x,y,z
 \right)  \right) ^{2}
=0.
\end{equation}

\section{ Method of solutions}

      To obtain  particular solutions of  nonlinear
       partial differential equations
\begin{equation}\label{Dr10}
F(x,y,f_x,f_y,f_{xx},f_{xy},f_{yy},f_{xxx},f_{xyy},f_{xxy},..)=0
 \end{equation}
    we use the parametric presentation of the functions and variables \cite{drum},\cite{dryum},\cite{cher}
\begin{equation}\label{Dr11}
f(x,y)\rightarrow u(x,t),\quad y \rightarrow v(x,t),\quad
f_x\rightarrow u_x-\frac{u_t}{v_t}v_x,\quad f_y \rightarrow \frac{u_t}{v_t},$$$$
\quad f_{yy} \rightarrow \frac{(\frac{u_t}{v_t})_t}{v_t}, \quad
f_{xy} \rightarrow \frac{(u_x-\frac{u_t}{v_t}v_x)_t}{v_t},...
\end{equation}
where variable $t$ is considered as parameter.

  Remark that conditions of equality of mixed derivatives
   \[
   f_{xy}=f_{yx}
   \]
are fulfilled at the such type of presentation.

  In result instead of equation (\ref{Dr10}) one gets the relation between a new
   variables $u(x,t)$ , $v(x,t)$ and their partial derivatives
\begin{equation}\label{Dr12}
\Psi(u,v,u_x,u_t,v_x,v_t...)=0.
  \end{equation}

    At the condition $v(x,t)=t$
  the relation (\ref{Dr12}) coincides with the equation (\ref{Dr10}) and takes a more general
form after the reduction  $u(x,t)=F(\omega(x,t),\omega_t...)$    and $v(x,t)=\Phi(\omega(x,t),\omega_t...)$.

    The substitution $u(x,t)$ into the relation (\ref{Dr12}) leads to the p.d.e. with respect
the function $v(x,t)$ and it can be considered as the partner equation to the equation (\ref{Dr10}).

      Classification of possible reductions of the relation (\ref{Dr12}) connected with a given equation
(\ref{Dr10}) has an important interest for development of the $(u,v)$-transformation method.

   The most popular reductions of the relation (\ref{Dr12}) are in the form
\[
u(x,t)={\frac {\partial }{\partial t}}\omega(x,t,z)t,
\]
\[
v(x,t)=t{\frac {\partial }{\partial t}}\omega(x,t,z)-\omega(x,t,z),
\]
or
\[
v(x,t)={\frac {\partial }{\partial t}}\omega(x,t,z)t,
\]
\[
u(x,t)=t{\frac {\partial }{\partial t}}\omega(x,t,z)-\omega(x,t,z).
\]

\section{An examples of solutions}

     The equation  (\ref{dr:eq2}) after the applying transformation
(\ref{Dr11}) with the conditions
\[
    v(x,t,z)=t{\frac {\partial }{\partial t}}\omega(x,t,z)-\omega(x,t,z)
\]
and
\[
u(x,t,z)={\frac {\partial }{\partial t}}\omega(x,t,z)
\]
  is reduced to the form
\begin{equation}\label{Dr14}
 \left( {\frac {\partial ^{2}}{\partial {t}^{2}}}\omega \left( x,t,z
 \right)  \right)  \left( {\frac {\partial ^{2}}{\partial x\partial z}
}\omega \left( x,t,z \right)  \right) ^{2}-2\, \left( {\frac {
\partial ^{2}}{\partial t\partial x}}\omega \left( x,t,z \right)
 \right)  \left( {\frac {\partial ^{2}}{\partial t\partial z}}\omega
 \left( x,t,z \right)  \right) {\frac {\partial ^{2}}{\partial x
\partial z}}\omega \left( x,t,z \right) +$$$$+ \left( {\frac {\partial ^{2}
}{\partial t\partial z}}\omega \left( x,t,z \right)  \right) ^{2}{
\frac {\partial ^{2}}{\partial {x}^{2}}}\omega \left( x,t,z \right) -
 \left( {\frac {\partial ^{2}}{\partial {t}^{2}}}\omega \left( x,t,z
 \right)  \right)  \left( {\frac {\partial ^{2}}{\partial {x}^{2}}}
\omega \left( x,t,z \right)  \right) {\frac {\partial ^{2}}{\partial {
z}^{2}}}\omega \left( x,t,z \right) + $$$$+\left( {\frac {\partial ^{2}}{
\partial t\partial x}}\omega \left( x,t,z \right)  \right) ^{2}{\frac
{\partial ^{2}}{\partial {z}^{2}}}\omega \left( x,t,z \right)
=0,
\end{equation}
which is equivalent the condition
\begin{equation}\label{Dr15}
 \left[ \begin {array}{ccc} {\frac {\partial ^{2}}{\partial {x}^{2}}}
\omega \left( x,t,z \right) &{\frac {\partial ^{2}}{\partial t
\partial x}}\omega \left( x,t,z \right) &{\frac {\partial ^{2}}{
\partial x\partial z}}\omega \left( x,t,z \right) \\\noalign{\medskip}
{\frac {\partial ^{2}}{\partial t\partial x}}\omega \left( x,t,z
 \right) &{\frac {\partial ^{2}}{\partial {t}^{2}}}\omega \left( x,t,z
 \right) &{\frac {\partial ^{2}}{\partial t\partial z}}\omega \left( x
,t,z \right) \\\noalign{\medskip}{\frac {\partial ^{2}}{\partial x
\partial z}}\omega \left( x,t,z \right) &{\frac {\partial ^{2}}{
\partial t\partial z}}\omega \left( x,t,z \right) &{\frac {\partial ^{
2}}{\partial {z}^{2}}}\omega \left( x,t,z \right) \end {array}
 \right] =0.
\end{equation}

    From solutions of the equation (\ref{Dr15}) can be derived solutions of the equation
       (\ref{dr:eq2}) with the help of elimination of the parameter $t$ from the relations
\[
    y-t{\frac {\partial }{\partial t}}\omega(x,t,z)+\omega(x,t,z)=0
\]
and
\[
F(x,y,z)-{\frac {\partial }{\partial t}}\omega(x,t,z)=0.
\]

    To integrate the equation (\ref{Dr15}) we rewrite it in the form
\begin{equation}\label{Dr16}
 \left( {\frac {\partial ^{2}}{\partial {y}^{2}}}h \left( x,y,z
 \right)  \right)  \left( {\frac {\partial ^{2}}{\partial x\partial z}
}h \left( x,y,z \right)  \right) ^{2}-2\, \left( {\frac {\partial ^{2}
}{\partial x\partial y}}h \left( x,y,z \right)  \right)  \left( {
\frac {\partial ^{2}}{\partial y\partial z}}h \left( x,y,z \right)
 \right) {\frac {\partial ^{2}}{\partial x\partial z}}h \left( x,y,z
 \right) + $$$$+\left( {\frac {\partial ^{2}}{\partial y\partial z}}h
 \left( x,y,z \right)  \right) ^{2}{\frac {\partial ^{2}}{\partial {x}
^{2}}}h \left( x,y,z \right) - \left( {\frac {\partial ^{2}}{\partial
{y}^{2}}}h \left( x,y,z \right)  \right)  \left( {\frac {\partial ^{2}
}{\partial {x}^{2}}}h \left( x,y,z \right)  \right) {\frac {\partial ^
{2}}{\partial {z}^{2}}}h \left( x,y,z \right) +$$$$+ \left( {\frac {
\partial ^{2}}{\partial x\partial y}}h \left( x,y,z \right)  \right) ^
{2}{\frac {\partial ^{2}}{\partial {z}^{2}}}h \left( x,y,z \right)
=0
\end{equation}
where we change the parameter $t$ on the variable $y$
and the function $\omega(x,t,z)$ on the function $h(x,y,z)$.

   After the $(u,v)$-transformation with conditions
\[
v \left( x,t,z \right) =t{\frac {\partial }{\partial t}}\theta \left(
x,t,z \right) -\theta \left( x,t,z \right),
 \]
\[
u \left( x,t,z \right) ={\frac {\partial }{\partial t}}\theta \left( x
,t,z \right)
\]
 this equation is reduced at the equation on the function $\theta(x,t,z)$
\begin{equation}\label{Dr17}
 \left( {\frac {\partial ^{2}}{\partial {x}^{2}}}\theta \left( x,t,z
 \right)  \right) {\frac {\partial ^{2}}{\partial {z}^{2}}}\theta
 \left( x,t,z \right) - \left( {\frac {\partial ^{2}}{\partial x
\partial z}}\theta \left( x,t,z \right)  \right) ^{2}=0.
\end{equation}

    From solutions of the equation (\ref{Dr17}) we find the function $h(x,y,z)$
by the way of elimination of the parameter $t$ from the relations
\[
y-t{\frac {\partial }{\partial t}}\theta \left(
x,t,z \right) +\theta \left( x,t,z \right)=0,
\]
and
\[
h(x,y,z)-{\frac {\partial }{\partial t}}\theta \left( x
,t,z \right)=0.
\]

     Using the function $h(x,y,z)$ we get the function $\omega(x,t,z) =h(x,t,z)$
and then the solutions of the equation (\ref{dr:eq2}).

    Solutions  of the equation (\ref{Dr17}) can be derived by the $(u,v)$ transformation
and are determined with the help of elimination of the parameter $\tau$ from the relations
\[
\tau x+\phi(\tau,t)z+\psi(\tau,t) \theta(x,t,z)-1=0,
\]
\[
x+\phi_{\tau} z+\psi_{\tau} \theta(x,t,z)=0
\]
where $\phi$ and $\psi$ are arbitrary functions.

\section{An example}

    After the substitution
\[
\varphi  \left( \tau,t \right) =-A \left( t \right) {\tau}^{2},
\]
\[
\psi \left( \tau,t \right) =B \left( t \right) \tau
\]
from the system of equations
\[
x\tau-A \left( t \right) {\tau}^{2}z+B \left( t \right) \tau\,\theta
 \left( x,t,z \right) -1=0,\]\[
x-2\,A \left( t \right) \tau\,z+B \left( t \right) \theta \left( x,t,z
 \right) =0
\]
we find
\[
\theta(x,t,z)={\frac {-x+2\,\sqrt {A \left( t \right) z}}{B \left( t \right) }}.
\]

    From the equations
\[
h \left( x,y,z \right) -{\frac {\partial }{\partial t}}\theta \left( x
,t,z \right)=0,\]
\[
y-{\frac {\partial }{\partial t}}\theta \left( x,t,z \right) +\theta
 \left( x,t,z \right)=0
\]
at the conditions
\[
A \left( t \right) = \left( B \left( t \right)  \right) ^{2},\]\[
B \left( t \right) ={\frac {t+1}{t-1}}
\]
we get the function
\[
h \left( x,y,z \right) =-{\frac { \left( 1/2\,\sqrt {2}x+1/2\,\sqrt {6
\,{x}^{2}-4\,yx+8\,\sqrt {z}x} \right) ^{2}}{x}}
\]
and corresponding function
\[
\omega \left( x,t,z \right) =-1/2\,{\frac { \left( x+\sqrt {x \left( 3
\,x-2\,t+4\,\sqrt {z} \right) } \right) ^{2}}{x}}.
\]

    Now after the elimination of the parameter $t$ from the system of equations
\[
F \left( x,y,z \right) -{\frac {\partial }{\partial t}}\omega \left( x
,t,z \right)=0\]\[
y-t{\frac {\partial }{\partial t}}\omega \left( x,t,z \right) +\omega
 \left( x,t,z \right)=0
\]
we find the function
\[
F \left( x,y,z \right) ={\frac {x+y+2\,\sqrt {z}+\sqrt {{y}^{2}-4\,yx-
4\,y\sqrt {z}+{x}^{2}+4\,\sqrt {z}x+4\,z}}{4\,\sqrt {z}+3\,x}}
\]
which is solution of the equation (\ref{dr:eq2}).

\section{Partner equation}

    After application of the $(u,v)$-transformation  with the condition
\[
u(x,t,z)=t
\]
the equation (\ref{Dr16})  is transformed to the partner equation
\begin{equation}\label{Dr18}
- \left( {\frac {\partial ^{2}}{\partial {t}^{2}}}v \left( x,t,z
 \right)  \right)  \left( {\frac {\partial ^{2}}{\partial x\partial z}
}v \left( x,t,z \right)  \right) ^{2}- \left( {\frac {\partial ^{2}}{
\partial t\partial z}}v \left( x,t,z \right)  \right) ^{2}{\frac {
\partial ^{2}}{\partial {x}^{2}}}v \left( x,t,z \right) - $$$$\left( {
\frac {\partial ^{2}}{\partial t\partial x}}v \left( x,t,z \right)
 \right) ^{2}{\frac {\partial ^{2}}{\partial {z}^{2}}}v \left( x,t,z
 \right) +2\, \left( {\frac {\partial ^{2}}{\partial t\partial x}}v
 \left( x,t,z \right)  \right)  \left( {\frac {\partial ^{2}}{
\partial t\partial z}}v \left( x,t,z \right)  \right) {\frac {
\partial ^{2}}{\partial x\partial z}}v \left( x,t,z \right) +$$$$+ \left( {
\frac {\partial ^{2}}{\partial {t}^{2}}}v \left( x,t,z \right)
 \right)  \left( {\frac {\partial ^{2}}{\partial {x}^{2}}}v \left( x,t
,z \right)  \right) {\frac {\partial ^{2}}{\partial {z}^{2}}}v \left(
x,t,z \right) =0.
\end{equation}

    After the substitution
\[
v \left( x,t,z \right) =A \left( x,t \right) z
\]
we get the equation with respect the function $A(x,t)$
\begin{equation}\label{Dr19}
 \left( {\frac {\partial ^{2}}{\partial {t}^{2}}}A \left( x,t \right)
 \right)  \left( {\frac {\partial }{\partial x}}A \left( x,t \right)
 \right) ^{2}+ \left( {\frac {\partial }{\partial t}}A \left( x,t
 \right)  \right) ^{2}{\frac {\partial ^{2}}{\partial {x}^{2}}}A
 \left( x,t \right) -\]\[-2\, \left( {\frac {\partial ^{2}}{\partial t
\partial x}}A \left( x,t \right)  \right)  \left( {\frac {\partial }{
\partial t}}A \left( x,t \right)  \right) {\frac {\partial }{\partial
x}}A \left( x,t \right) =0.
\end{equation}

    A simplest solution of this equation has the form
\[
A \left( x,t \right) ={\it \_F1} \left( x \right) {\it \_F2} \left( t
 \right),
\]
where the functions ${\it \_F1} \left( x \right)$ and ${\it \_F2} \left( x \right)$ are defined from the
equations
\[
{\frac {d^{2}}{d{x}^{2}}}{\it \_F1} \left( x \right) ={\frac { \left(
{\frac {d}{dx}}{\it \_F1} \left( x \right)  \right) ^{2}}{{\it \_c}_{{
1}}{\it \_F1} \left( x \right) }}=0,
\]
\[
{\frac {d^{2}}{d{t}^{2}}}{\it \_F2} \left( t \right) =2\,{\frac {
 \left( {\frac {d}{dt}}{\it \_F2} \left( t \right)  \right) ^{2}}{{
\it \_F2} \left( t \right) }}-{\frac { \left( {\frac {d}{dt}}{\it \_F2
} \left( t \right)  \right) ^{2}}{{\it \_F2} \left( t \right) {\it \_c
}_{{1}}}}=0
\]
and has the form
\[
{\it \_F1} \left( x \right) = \left(  \left( {\frac {{\it \_c}_{{1}}}{
-{\it \_C1}\,x+{\it \_C1}\,x{\it \_c}_{{1}}-{\it \_C2}+{\it \_C2}\,{
\it \_c}_{{1}}}} \right) ^{{\frac {{\it \_c}_{{1}}}{-1+{\it \_c}_{{1}}
}}} \right) ^{-1},
\]
\[
{\it \_F2} \left( t \right) = \left(  \left( {\frac {{\it \_C3}\,t-{
\it \_C3}\,t{\it \_c}_{{1}}+{\it \_C4}-{\it \_C4}\,{\it \_c}_{{1}}}{{
\it \_c}_{{1}}}} \right) ^{{\frac {{\it \_c}_{{1}}}{-1+{\it \_c}_{{1}}
}}} \right) ^{-1}.
\]

    In result we obtain the function $\omega \left( x,t,z \right)$
\[
\omega \left( x,t,z \right) =z \left(  \left( -{\frac {-{\it \_C3}\,y+
{\it \_C3}\,y{\it \_c}_{{1}}-{\it \_C4}+{\it \_C4}\,{\it \_c}_{{1}}}{{
\it \_c}_{{1}}}} \right) ^{{\frac {{\it \_c}_{{1}}}{-1+{\it \_c}_{{1}}
}}}\right) ^{-1}\times\]\[\times \left(  \left( {\frac {{\it \_c}_{{1}}}{-{\it \_C1}
\,x+{\it \_C1}\,x{\it \_c}_{{1}}-{\it \_C2}+{\it \_C2}\,{\it \_c}_{{1}
}}} \right) ^{{\frac {{\it \_c}_{{1}}}{-1+{\it \_c}_{{1}}}}} \right) ^
{-1}.
\]

  With the help of the $\omega \left( x,t,z \right)$ the function $F(x,y,z)$ can be fond from the relations
\[
F \left( x,y,z \right) -{\frac {\partial }{\partial t}}\omega \left( x
,t,z \right) =0
\]
and
\[
y-t{\frac {\partial }{\partial t}}\omega \left( x,t,z \right) +\omega
 \left( x,t,z \right) =0.
\]
after elimination of the parameter $t$.

    As example in the case
\[
{\it \_c}_{{1}}=2,
\]
the function $F(x,y,z)$ which is solution of the equation (\ref{dr:eq2}) is defined by the equation
\[
4\, \left( F \left( x,y,z \right)  \right) ^{3}{{\it \_C4}}^{3}+27\,
 \left( F \left( x,y,z \right)  \right) ^{2}{\it \_C3}\,z{{\it \_C1}}^
{2}{x}^{2}+54\, \left( F \left( x,y,z \right)  \right) ^{2}{\it \_C3}
\,z{\it \_C1}\,x{\it \_C2}+\]\[+27\, \left( F \left( x,y,z \right)
 \right) ^{2}{\it \_C3}\,z{{\it \_C2}}^{2}+12\, \left( F \left( x,y,z
 \right)  \right) ^{2}{\it \_C3}\,{{\it \_C4}}^{2}y+12\,F \left( x,y,z
 \right) {{\it \_C3}}^{2}{\it \_C4}\,{y}^{2}+\]\[+4\,{{\it \_C3}}^{3}{y}^{3
}=0.
\]

  The equation (\ref{Dr19}) can be integrated by the $(u,v)$- or the Legendre -transformation
and its solutions may be used to construction of solutions of the equation  (\ref{dr:eq2}).

\section{Ruled surfaces}

     The equation of any ruled $f=f(x,y)$ surface is derived by elimination of the parameter $\tau$ from
the relations
\[
f-\alpha(\tau) x-a(\tau)=0,
\]
\[
y-\beta(\tau) x-b(\tau)=0.
\]

    It can be presented as  \cite{git}
\[
 \left( {\frac {\partial ^{2}}{\partial {x}^{2}}}f \left( x,y \right)
 \right)  \left( {\frac {\partial }{\partial y}}Q \left( x,y \right)
 \right) ^{2}-2\, \left( {\frac {\partial ^{2}}{\partial x\partial y}}
f \left( x,y \right)  \right)  \left( {\frac {\partial }{\partial x}}Q
 \left( x,y \right)  \right) {\frac {\partial }{\partial y}}Q \left( x
,y \right) + $$$$+\left( {\frac {\partial ^{2}}{\partial {y}^{2}}}f \left(
x,y \right)  \right)  \left( {\frac {\partial }{\partial x}}Q \left( x
,y \right)  \right) ^{2}-8\,QT=0
\]
where
\[
T= \det\left[ \begin {array}{ccc} {\frac {\partial ^{2}}{\partial {x}^{2}}}f
 \left( x,y \right) &{\frac {\partial ^{2}}{\partial x\partial y}}f
 \left( x,y \right) &{\frac {\partial ^{2}}{\partial {y}^{2}}}f
 \left( x,y \right) \\\noalign{\medskip}{\frac {\partial ^{3}}{
\partial {x}^{3}}}f \left( x,y \right) &{\frac {\partial ^{3}}{
\partial {x}^{2}\partial y}}f \left( x,y \right) &{\frac {\partial ^{3
}}{\partial y\partial x\partial y}}f \left( x,y \right)
\\\noalign{\medskip}{\frac {\partial ^{3}}{\partial {x}^{2}\partial y}
}f \left( x,y \right) &{\frac {\partial ^{3}}{\partial y\partial x
\partial y}}f \left( x,y \right) &{\frac {\partial ^{3}}{\partial {y}^
{3}}}f \left( x,y \right) \end {array} \right],
\]
and
\[
Q=\left( {\frac {\partial ^{2}}{\partial {x}^{2}}}f \left( x,y \right)
 \right) {\frac {\partial ^{2}}{\partial {y}^{2}}}f \left( x,y
 \right) - \left( {\frac {\partial ^{2}}{\partial x\partial y}}f
 \left( x,y \right)  \right) ^{2},
\]

   In explicit form it looks as
\begin{equation}\label{Dr20}
-18\, \left( {\frac {\partial ^{2}}{\partial {x}^{2}}}f \left( x,y
 \right)  \right)  \left( {\frac {\partial ^{2}}{\partial {y}^{2}}}f
 \left( x,y \right)  \right)  \left( {\frac {\partial ^{3}}{\partial {
x}^{2}\partial y}}f \left( x,y \right)  \right)  \left( {\frac {
\partial ^{2}}{\partial x\partial y}}f \left( x,y \right)  \right) {
\frac {\partial ^{3}}{\partial y\partial x\partial y}}f \left( x,y
 \right) +$$$$+6\, \left( {\frac {\partial ^{2}}{\partial x\partial y}}f
 \left( x,y \right)  \right)  \left( {\frac {\partial ^{3}}{\partial {
x}^{3}}}f \left( x,y \right)  \right)  \left( {\frac {\partial ^{2}}{
\partial {y}^{2}}}f \left( x,y \right)  \right)  \left( {\frac {
\partial ^{2}}{\partial {x}^{2}}}f \left( x,y \right)  \right) {\frac
{\partial ^{3}}{\partial {y}^{3}}}f \left( x,y \right) -$$$$-6\, \left( {
\frac {\partial ^{2}}{\partial {y}^{2}}}f \left( x,y \right)  \right)
 \left( {\frac {\partial ^{3}}{\partial {x}^{2}\partial y}}f \left( x,
y \right)  \right)  \left( {\frac {\partial ^{2}}{\partial {x}^{2}}}f
 \left( x,y \right)  \right) ^{2}{\frac {\partial ^{3}}{\partial {y}^{
3}}}f \left( x,y \right) -$$$$-6\, \left( {\frac {\partial ^{2}}{\partial {
x}^{2}}}f \left( x,y \right)  \right) ^{2} \left( {\frac {\partial ^{3
}}{\partial {y}^{3}}}f \left( x,y \right)  \right)  \left( {\frac {
\partial ^{2}}{\partial x\partial y}}f \left( x,y \right)  \right) {
\frac {\partial ^{3}}{\partial y\partial x\partial y}}f \left( x,y
 \right) +$$$$+12\, \left( {\frac {\partial ^{3}}{\partial {x}^{3}}}f
 \left( x,y \right)  \right)  \left( {\frac {\partial ^{2}}{\partial {
y}^{2}}}f \left( x,y \right)  \right)  \left( {\frac {\partial ^{2}}{
\partial x\partial y}}f \left( x,y \right)  \right) ^{2}{\frac {
\partial ^{3}}{\partial y\partial x\partial y}}f \left( x,y \right) +$$$$+
12\, \left( {\frac {\partial ^{2}}{\partial x\partial y}}f \left( x,y
 \right)  \right) ^{2} \left( {\frac {\partial ^{3}}{\partial {x}^{2}
\partial y}}f \left( x,y \right)  \right)  \left( {\frac {\partial ^{2
}}{\partial {x}^{2}}}f \left( x,y \right)  \right) {\frac {\partial ^{
3}}{\partial {y}^{3}}}f \left( x,y \right) -$$$$-6\, \left( {\frac {
\partial ^{2}}{\partial x\partial y}}f \left( x,y \right)  \right)
 \left( {\frac {\partial ^{3}}{\partial {x}^{3}}}f \left( x,y \right)
 \right)  \left( {\frac {\partial ^{2}}{\partial {y}^{2}}}f \left( x,y
 \right)  \right) ^{2}{\frac {\partial ^{3}}{\partial {x}^{2}\partial
y}}f \left( x,y \right) -$$$$-6\, \left( {\frac {\partial ^{3}}{\partial {x
}^{3}}}f \left( x,y \right)  \right)  \left( {\frac {\partial ^{2}}{
\partial {y}^{2}}}f \left( x,y \right)  \right) ^{2} \left( {\frac {
\partial ^{2}}{\partial {x}^{2}}}f \left( x,y \right)  \right) {\frac
{\partial ^{3}}{\partial y\partial x\partial y}}f \left( x,y \right) +
 $$$$+\left( {\frac {\partial ^{2}}{\partial {x}^{2}}}f \left( x,y \right)
 \right) ^{3} \left( {\frac {\partial ^{3}}{\partial {y}^{3}}}f
 \left( x,y \right)  \right) ^{2}+ \left( {\frac {\partial ^{3}}{
\partial {x}^{3}}}f \left( x,y \right)  \right) ^{2} \left( {\frac {
\partial ^{2}}{\partial {y}^{2}}}f \left( x,y \right)  \right) ^{3}+$$$$+9
\, \left( {\frac {\partial ^{2}}{\partial {x}^{2}}}f \left( x,y
 \right)  \right)  \left( {\frac {\partial ^{2}}{\partial {y}^{2}}}f
 \left( x,y \right)  \right) ^{2} \left( {\frac {\partial ^{3}}{
\partial {x}^{2}\partial y}}f \left( x,y \right)  \right) ^{2}+$$$$+9\,
 \left( {\frac {\partial ^{2}}{\partial {y}^{2}}}f \left( x,y \right)
 \right)  \left( {\frac {\partial ^{2}}{\partial {x}^{2}}}f \left( x,y
 \right)  \right) ^{2} \left( {\frac {\partial ^{3}}{\partial y
\partial x\partial y}}f \left( x,y \right)  \right) ^{2}-$$$$-8\, \left( {
\frac {\partial ^{2}}{\partial x\partial y}}f \left( x,y \right)
 \right) ^{3} \left( {\frac {\partial ^{3}}{\partial {x}^{3}}}f
 \left( x,y \right)  \right) {\frac {\partial ^{3}}{\partial {y}^{3}}}
f \left( x,y \right)
=0
\end{equation}

   After the $(u,v)$-transformation
\[
u \left( x,t \right) =t{\frac {\partial }{\partial t}}\omega \left( x,
t \right) -\omega \left( x,t \right), \]\[
v \left( x,t \right) ={\frac {\partial }{\partial t}}\omega \left( x,t
 \right)
\]
we get the equation with respect the function $\omega \left( x,t \right) $
\begin{equation}\label{dr:eq6}
-9\, \left( {\frac {\partial ^{3}}{\partial x\partial t\partial x}}
\omega \left( x,t \right)  \right) ^{2} \left( {\frac {\partial ^{2}}{
\partial {x}^{2}}}\omega \left( x,t \right)  \right)  \left( {\frac {
\partial ^{2}}{\partial {t}^{2}}}\omega \left( x,t \right)  \right) ^{
2}-$$$$-6\, \left( {\frac {\partial ^{2}}{\partial {t}^{2}}}\omega \left( x
,t \right)  \right)  \left( {\frac {\partial ^{3}}{\partial x\partial
t\partial x}}\omega \left( x,t \right)  \right)  \left( {\frac {
\partial ^{2}}{\partial {x}^{2}}}\omega \left( x,t \right)  \right) ^{
2}{\frac {\partial ^{3}}{\partial {t}^{3}}}\omega \left( x,t \right) +$$$$+
6\, \left( {\frac {\partial ^{3}}{\partial {x}^{3}}}\omega \left( x,t
 \right)  \right)  \left( {\frac {\partial ^{2}}{\partial {t}^{2}}}
\omega \left( x,t \right)  \right) ^{2} \left( {\frac {\partial ^{2}}{
\partial {x}^{2}}}\omega \left( x,t \right)  \right) {\frac {\partial
^{3}}{\partial {t}^{2}\partial x}}\omega \left( x,t \right) +$$$$+ \left( {
\frac {\partial ^{3}}{\partial {x}^{3}}}\omega \left( x,t \right)
 \right) ^{2} \left( {\frac {\partial ^{2}}{\partial {t}^{2}}}\omega
 \left( x,t \right)  \right) ^{3}+9\, \left( {\frac {\partial ^{2}}{
\partial {t}^{2}}}\omega \left( x,t \right)  \right)  \left( {\frac {
\partial ^{2}}{\partial {x}^{2}}}\omega \left( x,t \right)  \right) ^{
2} \left( {\frac {\partial ^{3}}{\partial {t}^{2}\partial x}}\omega
 \left( x,t \right)  \right) ^{2}- $$$$-\left( {\frac {\partial ^{2}}{
\partial {x}^{2}}}\omega \left( x,t \right)  \right) ^{3} \left( {
\frac {\partial ^{3}}{\partial {t}^{3}}}\omega \left( x,t \right)
 \right) ^{2}=0
\end{equation}

 It has the solution of the form
$$
\omega \left( x,t \right) =A \left( t \right) +B \left( x \right) t.
$$
where the functions $A(t)$ and $B(t)$ satisfy the equations
\[
 \left( {\frac {d^{3}}{d{x}^{3}}}B \left( x \right)  \right) ^{2}-\mu
\, \left( {\frac {d^{2}}{d{x}^{2}}}B \left( x \right)  \right) ^{3}=0
\]
and
\[
9\, \left( {\frac {d^{2}}{d{t}^{2}}}A \left( t \right)  \right) ^{2}+6
\, \left( {\frac {d^{2}}{d{t}^{2}}}A \left( t \right)  \right) t{
\frac {d^{3}}{d{t}^{3}}}A \left( t \right) -\mu\,t \left( {\frac {d^{2
}}{d{t}^{2}}}A \left( t \right)  \right) ^{3}+{t}^{2} \left( {\frac {d
^{3}}{d{t}^{3}}}A \left( t \right)  \right) ^{2}=0.
\]

  From here we find
\[
A \left( t \right) = \left( -4\,{\frac {\ln  \left( {\it \_C1}\,t-1
 \right) }{\mu}}+4\,{\frac {\ln  \left( t \right) }{\mu}}-4\,{\frac {1
}{\mu\,{\it \_C1}\,t}}-{\frac {{\it \_C2}}{t}}+{\it \_C3} \right) t
\]
and
\[
B \left( t \right) =-4\,{\frac {\ln  \left( x+{\it \_C4} \right) }{\mu
}}+{\it \_C5}\,x+{\it \_C6}.
\]
    Using these expressions we find the function $f(x,y)$ satisfying the equation (\ref{Dr20}).

In particular case
\[
{\it \_C5}=0,\quad
{\it \_C2}=0,\quad
{\it \_C3}=0,\quad
{\it \_C4}=0,\quad
{\it \_C6}=0
\]
it is determined from the equation
\[
y\mu+8\,\ln  \left( 2 \right) +4\,\ln  \left( -{\frac {1}{f \left( x,y
 \right) \mu\,{\it \_C1}}} \right) -4\,\ln  \left( {\frac {-4+f
 \left( x,y \right) \mu\,{\it \_C1}}{f \left( x,y \right) \mu\,{{\it
\_C1}}^{2}}} \right) -\]\[-f \left( x,y \right) \mu\,{\it \_C1}+4\,\ln
 \left( x \right) =0
\]

\subsection{Partner equation}

        After the $(u,v)$-transformation with condition
\[
u \left( x,t \right) =t
\]
the equation (\ref{Dr20}) takes the form

\begin{equation}\label{dr:eq71}
-6\, \left( {\frac {\partial ^{2}}{\partial x\partial t}}v \left( x,t
 \right)  \right)  \left( {\frac {\partial ^{3}}{\partial {x}^{3}}}v
 \left( x,t \right)  \right)  \left( {\frac {\partial ^{2}}{\partial {
t}^{2}}}v \left( x,t \right)  \right)  \left( {\frac {\partial ^{2}}{
\partial {x}^{2}}}v \left( x,t \right)  \right) {\frac {\partial ^{3}}
{\partial {t}^{3}}}v \left( x,t \right) +$$$$+18\, \left( {\frac {\partial
^{2}}{\partial {x}^{2}}}v \left( x,t \right)  \right)  \left( {\frac {
\partial ^{2}}{\partial {t}^{2}}}v \left( x,t \right)  \right)
 \left( {\frac {\partial ^{3}}{\partial {x}^{2}\partial t}}v \left( x,
t \right)  \right)  \left( {\frac {\partial ^{2}}{\partial x\partial t
}}v \left( x,t \right)  \right) {\frac {\partial ^{3}}{\partial x
\partial {t}^{2}}}v \left( x,t \right) +$$$$+6\, \left( {\frac {\partial ^{
2}}{\partial x\partial t}}v \left( x,t \right)  \right)  \left( {
\frac {\partial ^{3}}{\partial {x}^{3}}}v \left( x,t \right)  \right)
 \left( {\frac {\partial ^{2}}{\partial {t}^{2}}}v \left( x,t \right)
 \right) ^{2}{\frac {\partial ^{3}}{\partial {x}^{2}\partial t}}v
 \left( x,t \right) +$$$$+6\, \left( {\frac {\partial ^{2}}{\partial {t}^{2
}}}v \left( x,t \right)  \right)  \left( {\frac {\partial ^{3}}{
\partial {x}^{2}\partial t}}v \left( x,t \right)  \right)  \left( {
\frac {\partial ^{2}}{\partial {x}^{2}}}v \left( x,t \right)  \right)
^{2}{\frac {\partial ^{3}}{\partial {t}^{3}}}v \left( x,t \right) -$$$$-12
\, \left( {\frac {\partial ^{2}}{\partial x\partial t}}v \left( x,t
 \right)  \right) ^{2} \left( {\frac {\partial ^{3}}{\partial {x}^{2}
\partial t}}v \left( x,t \right)  \right)  \left( {\frac {\partial ^{2
}}{\partial {x}^{2}}}v \left( x,t \right)  \right) {\frac {\partial ^{
3}}{\partial {t}^{3}}}v \left( x,t \right) +$$$$+6\, \left( {\frac {
\partial ^{3}}{\partial {x}^{3}}}v \left( x,t \right)  \right)
 \left( {\frac {\partial ^{2}}{\partial {t}^{2}}}v \left( x,t \right)
 \right) ^{2} \left( {\frac {\partial ^{2}}{\partial {x}^{2}}}v
 \left( x,t \right)  \right) {\frac {\partial ^{3}}{\partial x
\partial {t}^{2}}}v \left( x,t \right) +$$$$+6\, \left( {\frac {\partial ^{
2}}{\partial {x}^{2}}}v \left( x,t \right)  \right) ^{2} \left( {
\frac {\partial ^{3}}{\partial {t}^{3}}}v \left( x,t \right)  \right)
 \left( {\frac {\partial ^{2}}{\partial x\partial t}}v \left( x,t
 \right)  \right) {\frac {\partial ^{3}}{\partial x\partial {t}^{2}}}v
 \left( x,t \right) -$$$$-12\, \left( {\frac {\partial ^{3}}{\partial {x}^{
3}}}v \left( x,t \right)  \right)  \left( {\frac {\partial ^{2}}{
\partial {t}^{2}}}v \left( x,t \right)  \right)  \left( {\frac {
\partial ^{2}}{\partial x\partial t}}v \left( x,t \right)  \right) ^{2
}{\frac {\partial ^{3}}{\partial x\partial {t}^{2}}}v \left( x,t
 \right) +$$$$+8\, \left( {\frac {\partial ^{2}}{\partial x\partial t}}v
 \left( x,t \right)  \right) ^{3} \left( {\frac {\partial ^{3}}{
\partial {x}^{3}}}v \left( x,t \right)  \right) {\frac {\partial ^{3}}
{\partial {t}^{3}}}v \left( x,t \right) -9\, \left( {\frac {\partial ^
{2}}{\partial {t}^{2}}}v \left( x,t \right)  \right)  \left( {\frac {
\partial ^{2}}{\partial {x}^{2}}}v \left( x,t \right)  \right) ^{2}
 \left( {\frac {\partial ^{3}}{\partial x\partial {t}^{2}}}v \left( x,
t \right)  \right) ^{2}-$$$$-9\, \left( {\frac {\partial ^{2}}{\partial {x}
^{2}}}v \left( x,t \right)  \right)  \left( {\frac {\partial ^{2}}{
\partial {t}^{2}}}v \left( x,t \right)  \right) ^{2} \left( {\frac {
\partial ^{3}}{\partial {x}^{2}\partial t}}v \left( x,t \right)
 \right) ^{2}- \left( {\frac {\partial ^{2}}{\partial {x}^{2}}}v
 \left( x,t \right)  \right) ^{3} \left( {\frac {\partial ^{3}}{
\partial {t}^{3}}}v \left( x,t \right)  \right) ^{2}-$$$$- \left( {\frac {
\partial ^{3}}{\partial {x}^{3}}}v \left( x,t \right)  \right) ^{2}
 \left( {\frac {\partial ^{2}}{\partial {t}^{2}}}v \left( x,t \right)
 \right) ^{3}.
=0
\end{equation}

     It has solution of the form
     \[
     v(x, t) = A(t)+B(x)\]
which looks as
\[
v \left( x,t \right) =-4\,{\frac {\ln  \left( {\it \_C1}+x \right) }{
\mu}}+{\it \_C2}\,x+{\it \_C3}
4\,{\frac {\ln  \left( {\it \_C4}+t \right) }{\mu}}+{\it \_C5}\,t+{
\it \_C6}.
\]

     Using these expressions and the conditions
     \[
     y-v \left( x,t \right)=0,\quad t=f(x,y)
     \]
we can find the function $f(x,y$ which is solution of the equation (\ref{Dr20}.

In particular case
\[
{\it \_C6}=0,\quad
{\it \_C3}=0,\quad
{\it \_C1}=0,\quad
{\it \_C4}=0,\quad
{\it \_C5}=1
\]
one gets
\[
f\left( x,y \right) =4\,{\frac {{\it LambertW} \left( 1/4\,\mu\,{
{\rm e}^{1/4\,y\mu-1/4\,{\it \_C2}\,\mu\,x}}x \right) }{\mu}},
\]
where the function ${\it LambertW} \left( x \right)$ is defined  by the relation
\[
{\it LambertW} \left( x \right) {{\rm e}^{{\it LambertW} \left( x
 \right) }}=x.
\]

\section{Minimal surfaces}

     Minimal surfaces are defined by solutions of the equation \cite{Kur}
 \begin{equation}\label{Dr21}
\left( 1+ \left( {\frac {\partial }{\partial y}}f \left( x,y \right)
 \right) ^{2} \right) {\frac {\partial ^{2}}{\partial {x}^{2}}}f
 \left( x,y \right) -2\, \left( {\frac {\partial }{\partial y}}f
 \left( x,y \right)  \right)  \left( {\frac {\partial }{\partial x}}f
 \left( x,y \right)  \right) {\frac {\partial ^{2}}{\partial y
\partial x}}f \left( x,y \right) + \]\[+\left( 1+ \left( {\frac {\partial }
{\partial x}}f \left( x,y \right)  \right) ^{2} \right) {\frac {
\partial ^{2}}{\partial {y}^{2}}}f \left( x,y \right)=0
\end{equation}

      After the $(u,v)$-transformation
      \[
      v \left( x,t \right) =t{\frac {\partial }{\partial t}}\omega \left( x,
t \right) -\omega \left( x,t \right),\]\[
u \left( x,t \right) ={\frac {\partial }{\partial t}}\omega \left( x,t
 \right).
\]
one gets the equation
    \begin{equation}\label{Dr22}
- \left( {\frac {\partial ^{2}}{\partial x\partial t}}\omega \left( x,
t \right)  \right) ^{2}+{t}^{2} \left( {\frac {\partial ^{2}}{
\partial {t}^{2}}}\omega \left( x,t \right)  \right) {\frac {\partial
^{2}}{\partial {x}^{2}}}\omega \left( x,t \right) -{t}^{2} \left( {
\frac {\partial ^{2}}{\partial x\partial t}}\omega \left( x,t \right)
 \right) ^{2}+$$$$+2\,t \left( {\frac {\partial ^{2}}{\partial x\partial t}
}\omega \left( x,t \right)  \right) {\frac {\partial }{\partial x}}
\omega \left( x,t \right) -\left( {\frac {\partial }{\partial x}}
\omega \left( x,t \right)  \right) ^{2}-1+ \left( {\frac {\partial ^{2
}}{\partial {t}^{2}}}\omega \left( x,t \right)  \right) {\frac {
\partial ^{2}}{\partial {x}^{2}}}\omega \left( x,t \right)=0.
\end{equation}

     It has the solution
    \[
   \omega \left( x,t \right) =1/4\,t\arctan \left( t \right) + \left( 1+{
t}^{2} \right) {x}^{2}.
    \]

     Corresponding solution of the equation (\ref{Dr22}) can be presented in
a parametric form
\[
 f \left( x,y \right) =1/4\,{\frac {\arctan \left( t \right) +\arctan
 \left( t \right) {t}^{2}+t+8\,t{x}^{2}+8\,{t}^{3}{x}^{2}}{1+{t}^{2}}},
\]
\[
t=1/4\,{\frac {\sqrt {2}\sqrt {4\,y-1+\sqrt {16\,{y}^{2}-8\,y+1+64\,{x
}^{2}y+64\,{x}^{4}}}}{x}}.
\]

\section{Partner equation}

     After the $(u,v)$-transformation with the condition
      \[
      u \left( x,t \right) =t
\]
the equation (\ref{Dr21}) is transformed into the partner equation
  \begin{equation}\label{Dr23}
  - \left( {\frac {\partial }{\partial x}}v \left( x,t \right)  \right)
^{2}{\frac {\partial ^{2}}{\partial {t}^{2}}}v \left( x,t \right) +2\,
 \left( {\frac {\partial }{\partial x}}v \left( x,t \right)  \right)
 \left( {\frac {\partial ^{2}}{\partial x\partial t}}v \left( x,t
 \right)  \right) {\frac {\partial }{\partial t}}v \left( x,t \right)
-{\frac {\partial ^{2}}{\partial {t}^{2}}}v \left( x,t \right) -\]\[-{
\frac {\partial ^{2}}{\partial {x}^{2}}}v \left( x,t \right) - \left(
{\frac {\partial ^{2}}{\partial {x}^{2}}}v \left( x,t \right)
 \right)  \left( {\frac {\partial }{\partial t}}v \left( x,t \right)
 \right) ^{2}=0.
\end{equation}

    It has the same form with the initial equation (\ref{Dr21}).

     This property can be used to construction a new solutions of the equation
     (\ref{Dr21}) by the way of elimination of the parameter $t$ from the relations
 \begin{equation}\label{Dr24}
y-v\left(x,t\right)=0,\quad
t-f\left(x,y\right)=0.
\end{equation}

     Let us consider an examples.

     The function
     \[
     f \left( x,y \right) =\ln  \left( \sqrt {{x}^{2}+{y}^{2}}+\sqrt {{x}^{
2}+{y}^{2}-1} \right)
\]
 is solution of the equation (\ref{Dr21}).

     Then the function
     \[
     v \left( x,t \right) =\ln  \left( \sqrt {{x}^{2}+{t}^{2}}+\sqrt {{x}^{
2}+{t}^{2}-1} \right)
\]
is the  solution of the equation (\ref{Dr23}).

     Elimination of the parameter $t$ from the conditions
   \[
   y-\ln  \left( \sqrt {{x}^{2}+{t}^{2}}+\sqrt {{x}^{2}+{t}^{2}-1}
 \right)=0,
\]
 \[
 t=f\left(x,y\right)
 \]
 leads to the relation
 \[
 y-\ln  \left( \sqrt {{x}^{2}+ \left( f \left( x,y \right)  \right) ^{2
}}+\sqrt {{x}^{2}+ \left( f \left( x,y \right)  \right) ^{2}-1}
 \right)=0
\]
 from which we get the function
 \[
 f \left( x,y \right) =1/2\,\sqrt {2-4\,{x}^{2}+{{\rm e}^{-2\,y}}+{
{\rm e}^{2\,y}}}
\]
which is solution of the equation  (\ref{Dr21}).

     If the
     \[
     f \left( x,y \right) =\ln  \left( {\frac {\cos \left( y \right) }{\cos
 \left( x \right) }} \right)
\]
is solution of the equation (\ref{Dr21}),
then the function
\[
v \left( x,t \right) =\ln  \left( {\frac {\cos \left( t \right) }{\cos
 \left( x \right) }} \right)
\]
satisfies the equation (\ref{Dr23}).

    Now from the conditions (\ref{Dr24}) we find  the equation
    \[
    y-\ln  \left( {\frac {\cos \left( f \left( x,y \right)  \right) }{\cos
 \left( x \right) }} \right)=0
\]
from which we get a new solution of the equation (\ref{Dr21})
\[
f \left( x,y \right) =\arccos \left( {{\rm e}^{y}}\cos \left( x
 \right)  \right).
\]

    In the case of the substitution
     \[
v \left( x,t \right) =H \left( {x}^{2}+{t}^{2}+1 \right) )
\]
we find the solution of the equation (\ref{Dr23})
\[
v \left( x,t \right) =\ln  \left(  \left( -2\,{{\it \_C1}}^{2}+{x}^{2}
+{t}^{2}+\sqrt {-{\frac { \left( {x}^{2}+{t}^{2} \right)  \left( -{t}^
{2}+4\,{{\it \_C1}}^{2}-{x}^{2} \right) }{{{\it \_C1}}^{2}}}}{\it \_C1
} \right) {{\it \_C1}}^{-1} \right) {\it \_C1}.
\]

     Now from the conditions (\ref{Dr24}) we obtain the equation
\[
y=\]\[=\ln  \left(  \left( -2\,{{\it \_C1}}^{2}\!+\!{x}^{2}\!+\!f \left( x,
y \right)^{2}\!+\!\sqrt {{\frac { \left( {x}^{2}\!+\!f
 \left( x,y \right)^{2} \right)  \left( f \left( x,y
 \right)^{2}\!-\!4\,{{\it \_C1}}^{2}\!+\!{x}^{2} \right) }{{{\it
\_C1}}^{2}}}}{\it \_C1} \right) {{\it \_C1}}^{-1} \right) {\it \_C1}
\]
to determination of the function $f(x,y)$.

   Corresponding solution has the form
\[
f \left( x,y \right) =1/2\,\sqrt {2}\sqrt {{{\rm e}^{{\frac {y}{{\it
\_C1}}}}} \left( {\it \_C1}\,{{\rm e}^{2\,{\frac {y}{{\it \_C1}}}}}+4
\,{{\it \_C1}}^{2}{{\rm e}^{{\frac {y}{{\it \_C1}}}}}-2\,{{\rm e}^{{
\frac {y}{{\it \_C1}}}}}{x}^{2}+4\,{{\it \_C1}}^{3} \right) }{{\rm e}^
{-{\frac {y}{{\it \_C1}}}}}.
\]

\section{Acknowledgements}

The research was partially supported in the framework of joint Russian-Moldavian
research project\\[1mm]
(Grant 08.820.08.07 RF of HCSTD ASM, Moldova, and RFBR grant 08-01-90104, Russia).


\begin{thebibliography}{909}

\bibitem{Bysh} Byushgens S.S., Differentzialnaya geometriya, GTI T-TL, M.-L.,1940;

 \bibitem{drum} V. Dryuma, On theory of surfaces defined by the first order systems of equations,
{\it Buletinul Academiei de Stiinte a Repulicii Moldova, Matematica}, No.1(56), 2008, pp.161-175;

\bibitem{dryum} V.S. Dryuma, On solutions of a Heavenly equations and their generalization,
  {\it ArXiv:
  gr-qc/0611001}, v1 31 Oct 2006, p.1-13;

 \bibitem{cher} V. Dryuma, Multidimensional the Ricci-flat spaces defined by nonlinear equations,\\[1mm]
{\it arXiv:0911.2799 v1 [physics.gen-ph]}, 14 Nov 2009, pp.1-12;

 \bibitem{git} Gitomirsky O.K.,Lvovsky V.D.,Milinsky V.I., Zadachi po vysshei geometrii, v.2.,
 ONTI, M.-L.,1937;

\bibitem{Kur} Kurant R., Uravneiya s chastnymi proizvodnymi, MIR, M.,1964.
\end{thebibliography}
\end{document}